\documentclass[10pt]{iopart}

\usepackage{graphicx}  
\usepackage{dcolumn}
\usepackage{bm}

\begin{document}

\title[First-principles  study  of  Heusler  alloys
Mn$_2$VZ (Z=Al, Ge)] {First-principles  study  of exchange
interactions  and Curie temperatures of half-metallic
ferrimagnetic full Heusler  alloys Mn$_2$VZ (Z=Al, Ge)}

\author{E.~\c Sa\c s\i o\~glu\footnote[1]{Electronic mail: ersoy@mpi-halle.de} L. M.  Sandratskii, and  P. Bruno}

\address{Max-Planck Institut f\"ur Mikrostrukturphysik,
D-06120 Halle, Germany}


\begin{abstract}

We report the parameter-free, density functional theory
calculations of interatomic exchange interactions and Curie
temperatures of half-metallic ferrimagnetic full Heusler alloys
Mn$_2$VZ (Z=Al, Ge). To calculate the interatomic exchange
interactions we employ the   frozen-magnon approach. The Curie
temperatures are calculated within the mean--field approximation
to the classical Heisenberg Hamiltonian by solving a matrix
equation for a multi-sublattice system. Our calculations show
that, although a large magnetic moment is carried by Mn atoms,
competing ferromagnetic (inter sublattice) and antiferromagnetic
(intra sublattice) Mn-Mn interactions in Mn$_2$VAl almost cancel
each other in the mean-field experienced by the Mn atoms. In
Mn$_2$VGe the leading Mn-Mn exchange interaction is
antiferromagnetic. In both compounds the ferromagnetism of the Mn
subsystem is favored  by strong antiferromagnetic Mn-V
interactions. The obtained value of the Curie temperature of
Mn$_2$VAl is in good agrement with  experiment. For Mn$_2$VGe
there is no experimental information available and our calculation
is a prediction.

\end{abstract}



\section{Introduction}

The increased interest in the field of spin electronics attracts
strong attention to the magnetic systems suitable for the
realization of spin injection into a semiconductor \cite{ohno}.
The Heusler alloys are considered as a promising class of
materials that can possess the necessary combination of
properties. Among the features useful for applications are high
Curie temperature, high electron spin polarization at the Fermi
level and very small lattice mismatch with widely employed
semiconductors \cite{inas_3}. Some of the Heusler compounds were
found to have half-metallic ground state \cite{groot}
characterized by a 100\% spin-polarization of the charge carriers.
An interesting combination of physical properties makes Heusler
alloys the subject of intensive experimental and theoretical
investigations \cite{galana,exp_1,exp_2,ref_7,ref_9,ref_10}.

In full Heusler compounds characterized by the formula X$_2$YZ the
Mn atoms usually enter as the Y element. The compounds where Mn
assumes the X positions are very rare. Up to our knowledge there
are only two systems of this type studied experimentally:
Mn$_2$VAl \cite{itoh} and Mn$_2$VGa \cite{buschow}.

Mn$_2$VAl received much experimental and theoretical attention.
The neutron diffraction experiment by Itoh et al. \cite{itoh} gave
the ferrimagnetic state of compound with Mn magnetic moment of
$1.5\pm 0.3 \mu_B$ and V moment $-0.9 \mu_B$. Jiang {\textit et
al.} examined the magnetic structure of Mn$_2$VAl by X-ray
diffraction and magnetization measurements \cite{jiang}. They
found that Mn$_2$VAl was nearly half-metallic with the total
magnetic moment of 1.94$\mu_B$ at 5 K. The Curie temperature of
the sample was found to be about 760 K and the loss of
half-metallic character was attributed to the small amount of
disorder. The electron structure calculation by Ishida \textit {et
al.} performed within the local-density approximation (LDA) to the
density functional theory resulted in a ground state of Mn$_2$VAl
close to half-metallicity \cite{yoshida}. Recently a detailed
theoretical study of the magnetism of Mn$_2$VAl was reported by
Weht and  Pickett \cite{ruben} who used the generalized gradient
approximation (GGA) for the exchange correlation potential and
have shown that Mn$_2$VAl is a half-metallic ferrimagnet with the
atomic moments of 1.5$\mu_B$ and -0.9$\mu_B$ on Mn and V in very
good agreement with experiment. The Fermi level was found to lie
in the minority spin band.


The main purpose of the present work is a detailed study of the
exchange interactions in two half-metallic Mn$_2$VZ compounds:
Mn$_2$VAl and Mn$_2$VGe. (Mn$_2$VGe is not yet synthesized. Its
half-metallicity is predicted theoretically \cite{galana}.) Both
intra-sublattice and inter-sublattice exchange interactions are
calculated. We show that the pattern of exchange interactions in
these systems deviates from the physical picture that can be
expected on the basis of the experimental information available.
Indeed, the Mn-Mn distance of 2.96{\AA}({3.04\AA})  in Mn$_2$VAl
(Mn$_2$VGe) is substantially smaller than the Mn-Mn distance of
about {4\AA} in the X$_2$MnZ-type Heusler alloys \cite{B2}. (For
Mn$_2$VGe we use the interatomic distance of Mn$_2$VGa
\cite{buschow}). On the other hand, it is comparable with the
Mn-Mn distance in the antiferromagnetic fcc Mn (2.73\AA)
\cite{fccMn}. According to the Bethe-Slater curve \cite{bethe}
there are physical reasons to expect that smaller distances
between the 3d atoms stimulate the formation of the
antiferromagnetic structure whereas larger distances make the
ferromagnetic structure energetically preferable. Among the
Heusler alloys, a smaller distance between pairs of the Mn atoms
is obtained in the case of random occupation by Mn and Z atoms of
the Y and Z sublattices (see, e.g., the system with the B2-type
crystal structure  in Ref. \cite{B2}). Indeed, the experiment
gives for such systems the antiferromagnetic ordering \cite{B2}.
Therefore in the case of Mn$_2$VAl an antiferromagnetism of the
two Mn sublattices can be expected.

Our study shows, however, that the situation is more complex. The
nearest-neighbor (nn) Mn--Mn exchange interaction is found to be
ferromagnetic whereas the next nn Mn--Mn interaction is
antiferromagnetic. As a result the contributions of different Mn
atoms into the exchange field experienced by a given Mn atom
compensate strongly not giving substantial contribution into the
Curie temperature. The main role in the formation of the magnetic
structure and the magnetic transition temperature is played by the
strong Mn--V exchange interaction. We show that also in Mn$_2$VGe
the ferromagnetic ordering of the Mn subsystem is governed by the
Mn--V exchange interaction.

The paper is  organized as follows. In section II we present the
calculational approach. Section III contains the results of the
calculations and discussion.  Section IV gives the conclusions.

\section{Calculational Method}

The calculations are carried out with the augmented spherical
waves (ASW) method \cite{asw} within the generalized gradient
approximation (GGA) \cite{gga} for the  exchange--correlation
potential. We use the experimental lattice  parameter of Mn$_2$VAl
\cite{B2,lp}. The radii of all atomic spheres are chosen equal.

\begin{figure}[ht]
\begin{center}
\includegraphics[scale=0.32]{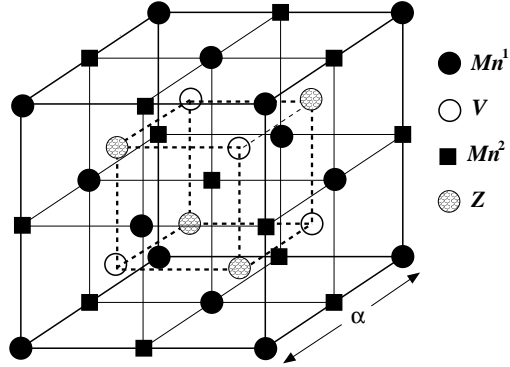}
\end{center}
\caption{ Schematic representation of the $L2_1$ structure. The
lattice consists of four interpenetrating fcc sublattices with the
positions $(0,0,0)$ and $(\frac{1}{2},\frac{1}{2},\frac{1}{2})$
for the Mn and $(\frac{1}{4},\frac{1}{4},\frac{1}{4})$ and
$(\frac{3}{4},\frac{3}{4},\frac{3}{4})$ for the V and Z,
respectively.}
 \label{crystal_structure}
\end{figure}

We describe the interatomic exchange interactions in terms of the
classical Heisenberg Hamiltonian
\begin{equation}
\label{eq:hamiltonian2}
 H_{eff}=-  \sum_{\mu,\nu}\sum_{\begin {array}{c}
^{{\bf R},{\bf R'}}\\ ^{(\mu{\bf R} \ne \nu{\bf R'})}\\
\end{array}} J_{{\bf R}{\bf R'}}^{\mu\nu}
{\bf s}_{\bf R}^{\mu}{\bf s}_{\bf R'}^{\nu}
\end{equation}
In Eq.(\ref{eq:hamiltonian2}), the  indices  $\mu$ and $\nu$
label different sublattices and ${\bf R}$ and ${\bf R'}$ are the
lattice vectors specifying the atoms within sublattices, ${\bf
s}_{\bf R}^\mu$ is the unit vector pointing in the direction of
the magnetic moment at site $(\mu,{\bf R})$. The systems
considered contain three 3{\textit d} atoms in the unit cell (see
Fig.\ref{crystal_structure}).

We employ the frozen--magnon approach \cite{magnon_2,magnon_3} to
calculate interatomic Heisenberg exchange parameters. The
calculations involve few steps. In the first step, the exchange
parameters between the atoms of a given sublattice $\mu$ are
computed. The calculation is based on the evaluation of the energy
of the frozen--magnon configurations defined by the following
atomic polar and azimuthal angles
\begin{equation}
\theta_{\bf R}^{\mu}=\theta, \:\: \phi_{\bf R}^{\mu}={\bf q \cdot
R}+\phi^{\mu}. \label{eq_magnon}
\end{equation}
The constant phase $\phi^{\mu}$ is always chosen equal to zero.
The magnetic moments of all other sublattices are kept parallel to
the z axis. Within the Heisenberg model~(\ref{eq:hamiltonian2})
the energy of such configuration takes the form
\begin{equation}
\label{eq:e_of_q} E^{\mu\mu}(\theta,{\bf
q})=E_0^{\mu\mu}(\theta)+\sin^{2}\theta J^{\mu\mu}({\bf q})
\end{equation}
where $E_0^{\mu\mu}$ does not depend on {\bf q} and the Fourier transform $J^{\mu\nu}({\bf q})$
is defined by
\begin{equation}
\label{eq:J_q}
J^{\mu\nu}({\bf q})=\sum_{\bf R}
J_{0{\bf R}}^{\mu\nu}\:\exp(i{\bf q\cdot R}).
\end{equation}

In the case of $\nu=\mu$ the sum in Eq. (\ref{eq:J_q}) does not
include ${\bf R}=0$. Calculating $ E^{\mu\mu}(\theta,{\bf q})$ for
a regular ${\bf q}$--mesh in the Brillouin zone of the crystal and
performing back Fourier transformation one gets exchange
parameters $J_{0{\bf R}}^{\mu\mu}$ for sublattice $\mu$.

The  determination of the exchange interactions between the atoms
of two different sublattices $\mu$ and $\nu$  is discussed in Ref.
\cite{intersublattice}.

\begin{table}
\caption{Magnetic moments (in $\mu_B$) of Mn$_2$VZ (Z= Al, Ge).
$a$ is the lattice parameter (in \AA).}
\begin{center}
\begin{tabular}{lllllll}
\\
\hline
  \hline
          $$ & $$ &     $a$(\AA)& Mn & V &  Z & Cell \\
 \hline
Mn$_2$VAl & Present work      & 5.932$^a$     & 1.525          & -1.022         & -0.031          &  2.00          \\
          & Weht$^b$          & 5.875         & 1.500          & -0.900         & -0.100          &  2.00          \\
          & Galanakis$^c$     & 5.932$^a$     & 1.413          & -0.786         & -0.020          &  2.02          \\
          & Jiang (Expt)$^d$  & 5.920         &   -            &  -             &  -              & \bf{1.94}     \\
Mn$_2$VGe & Present work      & 6.095$^e$     & 1.003          & -0.969         & -0.037          &  1.00          \\
          & Galanakis$^c$     & 5.932$^a$     & 0.750          & -0.476         & -0.020          &  1.00          \\
\hline
  \hline
\end{tabular}
\end{center}
$^a$Ref.\cite{B2}\\
$^b$Ref.\cite{ruben}\\
$^c$Ref.\cite{galana}\\
$^d$Ref.\cite{jiang}\\
$^e$Ref.\cite{lp}\\
\label{tab:moments}
\end{table}

The Curie  temperature is estimated  within the mean--field
approximation for a multi--sublattice material by solving the
system of  coupled   equations \cite{intersublattice,Anderson}
\begin{equation}
\label{eq_system} \langle s^{\mu}\rangle
=\frac{2}{3k_BT}\sum_{\nu}J_0^{\mu\nu}\langle s^{\nu}\rangle
\end{equation}
where  $\langle s^{\nu}\rangle$ is the average $z$ component of
${\bf s}_{{\bf R}}^{\nu}$ and $J_0^{\mu\nu}\equiv\sum_{\bf R}
J_{0{\bf R}}^{\mu\nu}$. Eq. (\ref{eq_system}) can be represented
in  form of an eigenvalue matrix problem
\begin{equation}
\label{eq_eigenvalue} ({\bf \Theta}-T {\bf I}){\bf S}=0
\end{equation}
where $\Theta_{\mu\nu}=\frac{2}{3k_B}J_0^{\mu\nu}$, ${\bf I}$ is a
unit matrix and ${\bf S}$ is the vector of $\langle s^{\nu}\rangle
$. The largest eigenvalue of matrix $\Theta$ gives the value of
the Curie temperature. \cite{Anderson}

\section{Results and Discussion}

The crystal structure is presented in Fig.\ref{crystal_structure}. The two Mn sublattices
are equivalent. The nearest Mn atoms belong to two different sublattices.

In Table \ref{tab:moments}  we present calculated magnetic
moments. For comparison, the available experimental values of the
moments and the results of previous calculations are given. The
net magnetic moment per unit cell  is 2$\mu_B$ for Mn$_2$VAl and
1$\mu_B$ for Mn$_2$VGe. The magnetic alignment is ferrimagnetic in
both systems. The Mn moments are parallel and assume the values
close to 1.5 $\mu_B$ in Mn$_2$VAl and to 1 $\mu_B$ in Mn$_2$VGe.
The moment of V is close to $-1\mu_B$ in both systems. The values
of the moments are in agreement with the results of previous
calculations.

\begin{figure}[ht]
\begin{center}
\includegraphics[scale=0.48]{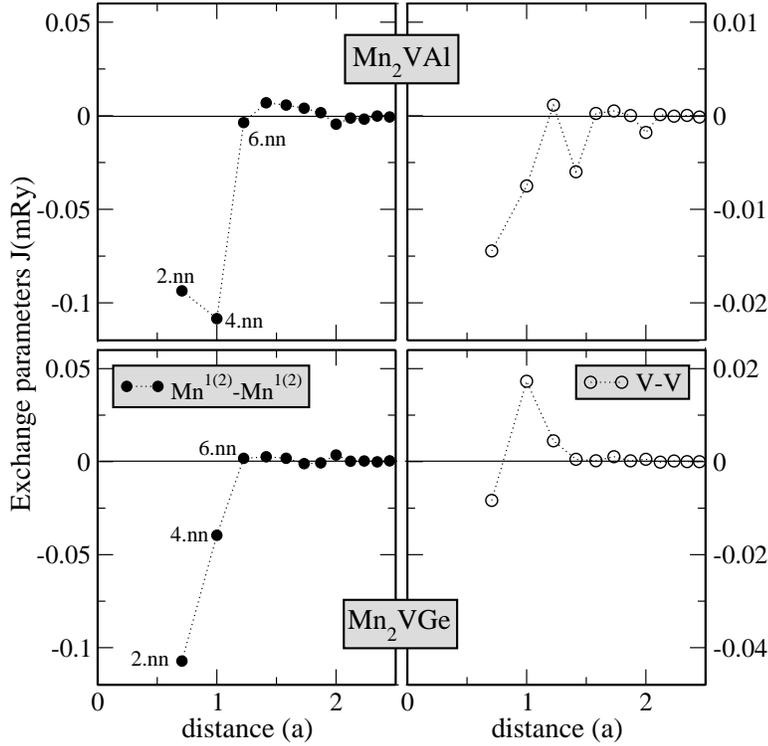}
\end{center}
\caption{Intra--sublattice Mn--Mn (left-hand part) and V--V
(right-hand part) exchange interactions  in Mn$_2$VZ (Z=Al and Ge)
as a function of the  distance  given in  units of the lattice
constant.}
 \label{exchange_intra}
\end{figure}

\begin{figure}[ht]
\begin{center}
\includegraphics[scale=0.48]{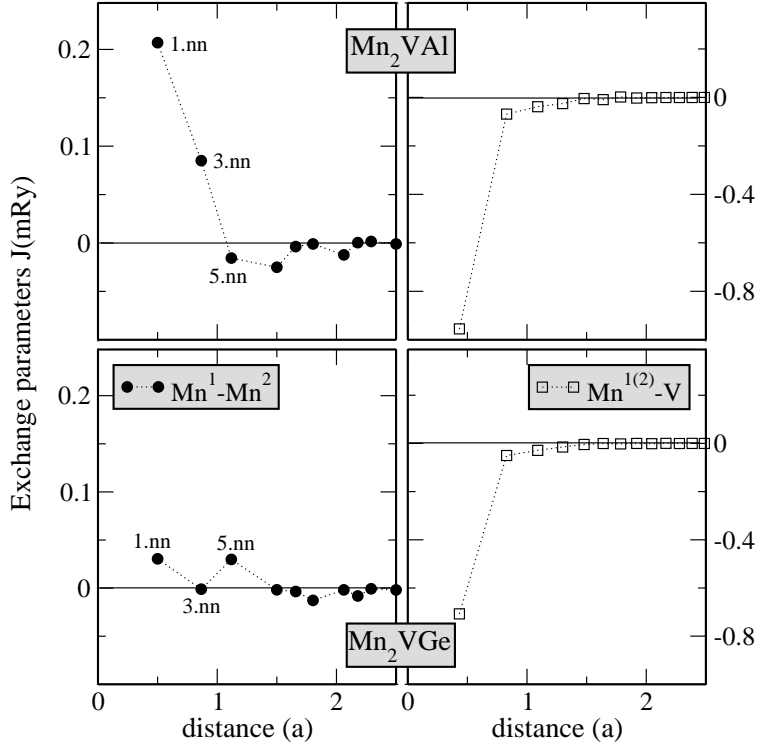}
\end{center}
\caption{Inter--sublattice Mn$^1$--Mn$^2$  (left side) and Mn--V
 (right side) exchange interactions  in
Mn$_2$VZ (Z=Al and  Ge) as a function of the  distance  given in
units of the lattice constant.}
 \label{exchange_inter}
\end{figure}

The calculated Heisenberg exchange parameters are presented in
figures~\ref{exchange_intra} and \ref{exchange_inter}. We obtained
a strong dependence  of the pattern of the  Mn--Mn and V--V
exchange interactions  on the type of the Z atom.
Similar result was obtained earlier
for Ni--based full Heusler alloys
\cite{intersublattice}. The nearest  Mn-Mn distance  is half
of the lattice constant $a$. The exchange interaction between the
nearest Mn atoms is ferromagnetic (Fig.~\ref{exchange_inter}).
Simultaneous examination of both intra- and inter-sublattice Mn-Mn
exchange interactions
(Figs.~\ref{exchange_intra},~\ref{exchange_inter}) indicates the
RKKY--type oscillations.

If only Mn-Mn exchange interactions are considered, in  Mn$_2$VAl
prevail ferromagnetic interactions while
in Mn$_2$VGe the Mn-Mn dominate antiferromagnetic interactions.
The corresponding Curie and  Ne\'{e}l
temperature are given in the second column of table~\ref{tab:Curietemperature}.

The interactions  between V atoms are very small and can be
neglected. The formation of the ferrimagnetic structure with all
Mn moments being parallel to each other and the V moments directed
oppositely is determined by the strong antiferromagnetic exchange
interactions between the nearest Mn and V moments. In Mn$_2$VAl
this interaction is  5 times  larger than the nearest neighbor
Mn--Mn interaction. In Mn$_2$VGe this factor increases to 20. The
strong Mn--V antiferromagnetic coupling makes a parallel direction
of the Mn moments surrounding a V atom energetically preferable
leading to the ferromagnetism of the Mn sublattices. In Mn$_2$VGe
this trend overcomes the direct antiferromagnetic Mn--Mn
interaction.

In the third column of Table~\ref{tab:Curietemperature} we present the
Curie temperature calculated with the Mn-Mn exchange interaction being
neglected. The Curie temperature given in the forth column takes into account
both Mn-V and Mn-Mn interactions. It is clearly seen that the main contribution
for both systems comes from the Mn-V interaction. In Mn$_2$VAl, the correction
of $T_c$ due to the Mn-Mn exchange interaction is positive and very small.
In Mn$_2$VGe, it is negative and amounts to 15\%.

In Mn$_2$VAl, where the experimental estimation of the Curie
temperature is available, both the theoretical and experimental
values are in good agreement. The theoretical Curie temperature of
Mn$_2$VGe should be considered as prediction.

\begin{table}
\caption{Mean--field estimation of the Curie temperatures  for
Mn$_2$VZ (Z$=$Al, Ge). The second (third) column gives the Curie
temperature calculated with account for Mn-Mn (Mn-V) interactions
only. In the fourth column both types of interactions are taken
into account. The experimental value of  Curie temperature for
Mn$_2$VAl is taken  from Ref.\cite{B2}.}
\label{tab:Curietemperature}
\begin{center}
\begin{tabular}{lcccc}
\\
\hline
  \hline
& $T_{c,N}^{{Mn-Mn}[MFA]}(K)$ &   $T_{c}^{{Mn-V}[MFA]}(K)$ &
$T_{c}^{[MFA]}(K)$ & $ T_{c}^{[Exp]}(K)$ \\
\hline
Mn$_2$VAl  & 30     &  623    &  638  & 760  \\
Mn$_2$VGe  & 170    &  488    &  413  & -    \\
\hline
  \hline
\end{tabular}
\end{center}
\end{table}

\section{Conclusion}

In conclusion, we have systematically studied exchange
interactions and Curie temperatures in two half--metallic
ferrimagnetic full Heusler alloys Mn$_2$VZ (Z= Al, Ge). The
calculations are performed within the parameter--free density
functional theory. We show that various Mn--Mn exchange
interactions compensate each other and the ferromagnetism of the
Mn subsystem is favored by very strong antiferromagnetic Mn--V
interactions. Good agreement with experiment is obtained for the
Curie temperature of Mn$_2$VAl. We give prediction for the Curie
temperature of  Mn$_2$VGe.

\ack The financial support of Bundesministerium f\"ur Bildung und
Forschung is acknowledged.


\begin{thebibliography}{}
\bibitem{lp}
To our knowledge, Mn$_2$VGe has not been studied experimentally.
We perform calculations of Mn$_2$VGe using the lattice  constants
of Mn$_2$VAl and  Mn$_2$VGa. In the lattice constant of Mn$_2$VGa
we found it half--metallic, while it was nearly half--metallic
with the lattice constant of Mn$_2$VAl. We will present the
results of calculation for the former case.
\end{thebibliography}
\end{document}